\documentclass[preprint2]{aastex}
\usepackage{rotating}

\shorttitle{DWARF ELLIPTICAL GALAXIES}
\shortauthors{Graham \& Guzm\'an}

\begin{document}

\title{{\it HST} PHOTOMETRY OF DWARF ELLIPTICAL GALAXIES IN COMA, 
AND AN EXPLANATION FOR THE ALLEGED STRUCTURAL DICHOTOMY BETWEEN DWARF 
AND BRIGHT ELLIPTICAL GALAXIES\altaffilmark{1}}

\author{Alister W.\ Graham and Rafael Guzm\'an}
\affil{Department of Astronomy, University of Florida, P.O.\ Box 112055, Gainesville, FL 32611, USA}
\email{Graham@astro.ufl.edu, Guzman@astro.ufl.edu}

\altaffiltext{1}{Based on observations made with the NASA/ESA {\sl
Hubble Space Telescope}, obtained at the Space Telescope Science
Institute, which is operated by the Association of Universities for
Research in Astronomy, Inc., under NASA contract NAS 5-26555.}

\begin{abstract}
As part of a research program exploring how and why dwarf 
elliptical (dE) galaxies depart from the Fundamental Plane defined 
by luminous elliptical (E) galaxies, we have analyzed archival 
{\it Hubble Space Telescope} F606W images of a sample of 18 
dE galaxy candidates in the Coma Cluster.  
We model the full radial extent of their light-profiles by 
simultaneously fitting a PSF-convolved S\'ersic $R^{1/n}$ 
model and, when necessary, either a central point-source or 
a central PSF-convolved Gaussian.  Nucleation was detected in 
all but two of our final sample of 15 dE galaxies. 
When detected, the luminosities of the central component $L_{\rm nuc}$ 
scale with the host galaxy luminosity $L_{\rm gal}$ such that 
$L_{\rm nuc} = 10^{4.76\pm 0.10}\left( L_{\rm gal}/10^7\right) ^{0.87\pm 0.26}$. 
We confirm that the light-profiles of the underlying host galaxies 
display systematic departures from an exponential model that are 
correlated with the model-independent host galaxy luminosity and 
are not due to biasing from the nuclear component.  
The Pearson correlation coefficient between $\log(n)$ and
central galaxy surface brightness $\mu_0$ (excluding the flux from 
extraneous central components) is -0.83 at a significance level of 99.99\%. 
Excluding one outlier, the Pearson correlation coefficient 
between the logarithm of the S\'ersic index $n$ and the host 
galaxy magnitude is -0.77 at a significance of 99.9\%.  
We explain the observed relationship between dE galaxy luminosity 
and the inner logarithmic profile slope $\gamma\prime$ as 
a by-product of the correlation between luminosity and S\'ersic index $n$. 
Including, from the literature, an additional 232 dE and E galaxies 
spanning 10 mag in absolute magnitude ($M$), the dE galaxies 
are shown to display a continuous sequence with the brighter E 
galaxies such that $\mu_0$ brightens linearly with $M$ until 
core formation causes the most luminous ($M_B\lesssim-20.5$ mag) 
E galaxies to deviate from this relation.  
The different 
behavior of dE and E galaxies in the $M$--$<$$\mu$$>$$_e$ 
(and $M$--$\mu_e$) diagram, and the $<$$\mu$$>$$_e$--$\log R_e$ 
diagram have nothing to do with core formation, and are in fact 
expected from the continuous and linear relation between 
$M$ and $\mu_0$, and $M$ and $\log(n)$. 
\end{abstract}


\section{Introduction}

The commonly accepted, albeit somewhat arbitrary, definition of whether 
an elliptical galaxy is a dwarf galaxy depends on whether its absolute 
magnitude is fainter than $M_B=-18$ (Sandage \& Binggeli 
1984).\footnote{A further distinction 
between faint and bright dwarf elliptical galaxies at 
$M_B=-16$ ($H_0=50$ km s$^{-1}$ Mpc$^{-1}$) has been used by 
Ferguson \& Binggeli (1994).}  
The realization that dwarf ellipticals could be reasonably well 
described with an 
exponential function (Faber \& Lin 1983; Binggeli, Sandage \& Tarenghi 1984) 
and that bright ellipticals are better fit with de Vaucouleurs' (1948, 1959) 
$R^{1/4}$-law led to the notion that they are two distinct families of 
galaxies (Wirth \& Gallagher 1984, but see also Graham 2002a).  
One of the most referenced papers to support this view is Kormendy (1985).  
Plotting central surface brightness against luminosity, 
Kormendy's Figure 3 shows two relations (almost at right angles to each 
other): one for the dE galaxies and the other for the luminous elliptical 
galaxies.  Similarly, diagrams using $\mu_e$, the surface brightness at the 
effective half-light radius $R_e$, or $<\mu>_e$, the average surface 
brightness within $R_e$, also show two somewhat perpendicular relations.  
The reason for these differences, generally interpreted as evidence for 
a different formation mechanism, are far from broadly understood 
and will therefore be addressed in this paper.

In contrast to the above, there is substantial evidence 
for a continuity, rather than a dichotomy, between the alleged elliptical 
(E) and dwarf elliptical (dE) galaxy classes. 
%
%
Caldwell (1983; their Figure 6) showed that, fainter than $M_B\sim -20.5$, 
there is a continuous 
trend between the central surface brightnesses and absolute magnitudes 
of dE and E galaxies --- more luminous galaxies have brighter central 
surface brightnesses 
(Caldwell 1987; Caldwell and Bothun 1987; Ferguson \& Sandage 1988; 
Karachentsev et al.\ 1995; Hilker et al.\ 1999; Drinkwater et al.\ 2001; 
Hilker, Mieske, \& Infante 2003).  This observation has 
been highlighted by Jerjen \& Binggeli (1997) and Jerjen, Binggeli, 
\& Freeman (2000a) note that even the brightest elliptical 
galaxies (with partially evacuated cores) follow the same continuous 
and linear trend if one uses the inwardly extrapolated 
central surface brightness of the best-fitting S\'ersic model (fitted 
outside of the core region). 
Caldwell (1983) also revealed a continuous and linear relationship 
exists between ($U-V$) color and luminosity over the magnitude interval 
$-23 < M_V < -15$ (see Terlevich, Caldwell, \& Bower 2001, and 
Odell, Schombert, \& Rakos 2002 for recent reviews).  Scodeggio (2001) 
showed this relation may, however, be far weaker than previously thought 
once one allows for color gradients and uses effective radii rather than 
fixed aperture sizes (see also the simulations of Cole et al.\ 2000). 
Relatedly, Caldwell \& Bothun (1987) revealed a continuous luminosity--metallicity 
relation across the alleged dE/E classes 
%
%
(Barazza \& Binggeli 2002). 
Previous studies of the 
luminosity-metallicity relation predominantly dealt with brighter 
galaxy samples: Faber 1973; Terlevich et al.\ 1981; Tonry \& Davis 
1981; Dressler 1984; Vader 1986. 
The relation between luminosity and velocity dispersion for E galaxies 
(Faber \& Jackson 1976, Tonry 1981) has also been 
shown to extend linearly to include the dE galaxies (e.g., Bender, 
Burstein, \& Faber 1992).  
There is additionally a continuous correlation between luminosity and 
globular cluster metallicity (Forbes et al.\ 1996).  None of these 
correlations with luminosity suggest evidence for a discontinuity 
at $M_B\sim -18$.  

To address the evidence for a dichotomy, we must turn to the issue
of galaxy structure. 
It was realized over a decade ago that the light-profile shapes 
of dwarf and bright elliptical galaxies display, respectively, 
luminosity-dependent departures from the exponential and $R^{1/4}$ 
models (e.g., Capaccioli 1984, 1987; Michard 1985; Schombert 1986; 
Caldwell \& Bothun 1987; Djorgovski \& Kormendy 1989; Binggeli \& 
Cameron 1991; Prugniel, Bica, \& Alloin 1992; James 1991, 1994).  
Indeed, the luminosity dependence of galaxy concentration/profile-shape 
was known by Baade (1944). 

\begin{figure}
\epsscale{0.75}
\plotone{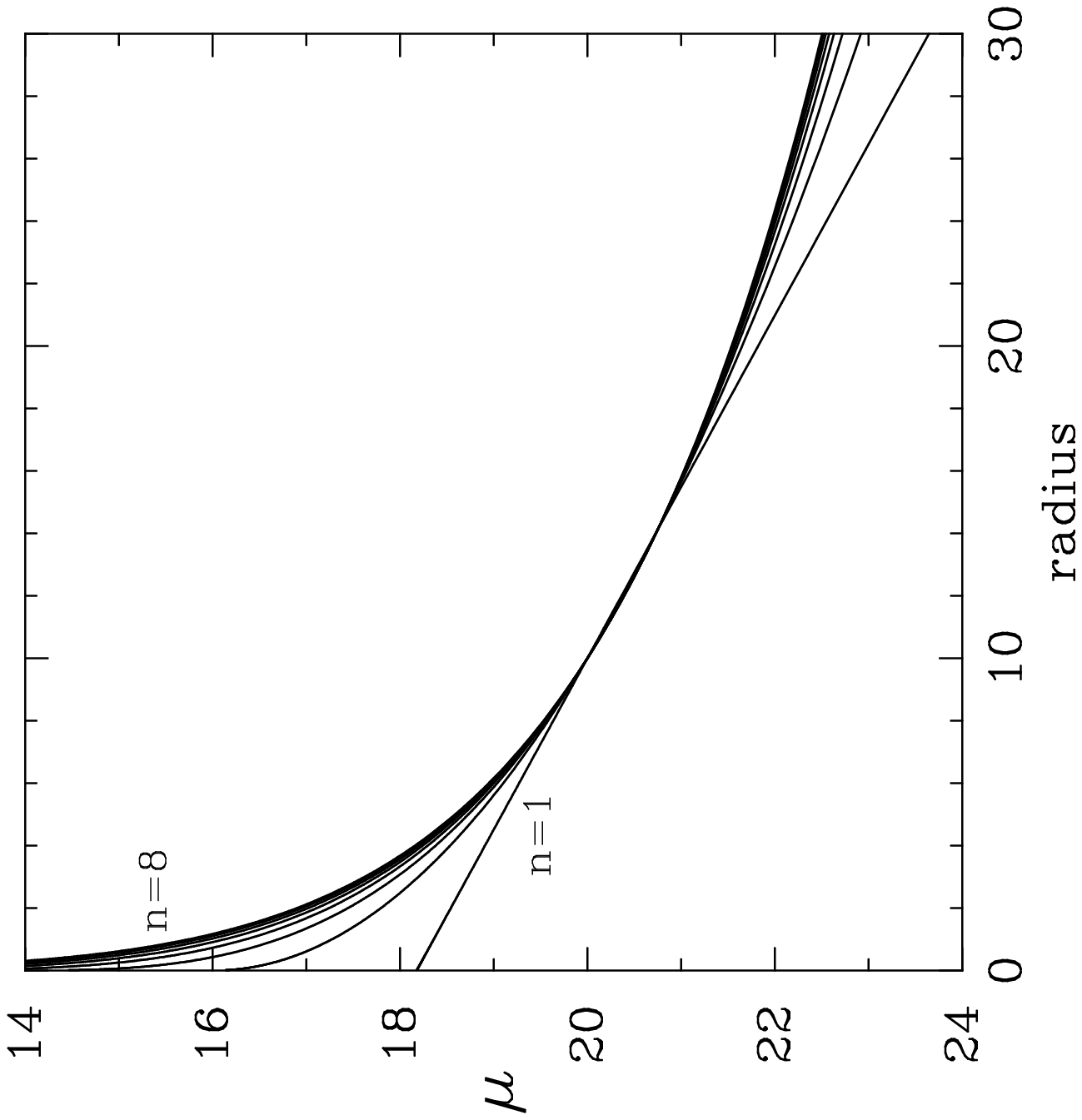}
\caption{Different $R^{1/n}$ S\'ersic models normalized at $R_e=10$ and $\mu_e=20$.}
\label{fig1}
\end{figure}

S\'ersic's (1968) $R^{1/n}$ model can encompass both de Vaucouleurs 
$R^{1/4}$-law and the exponential model, and a variety of other profile 
shapes by varying its exponent $n$.  It offers a means to explore the 
observed range of galaxy structures, and can be written as 
\begin{equation}
I(r)=I_e exp\left\{ -b\left[ \left( \frac{R}{R_e}\right) ^{1/n} -1\right] \right\}, 
\end{equation}
where the intensity $I$ at the half-light radius $R=R_e$ is denoted 
by $I_e$.  The third parameter $n$ describes the shape, or curvature, 
of the light-profile.  The quantity $b$ is a function of $n$, and 
is approximately $2n-1/3$ (Graham 2001).  
The differences between an $n=1$ and an $n=2$ (or 3) profile are quite 
distinct (see Figure~\ref{fig1}), and it is now almost universally accepted that dE galaxies 
possess different profile shapes.  
However, in the case of brighter elliptical galaxies, as noted by 
Makino, Akiyama, \& Sugimoto (1990), real departures from an $R^{1/4}$ 
model are often missed.  One reason for this is because of 
the similarity of the S\'ersic model when 
dealing with a restricted radial range of a galaxy's surface 
brightness profile and when $3\lesssim n \lesssim 10$. 
Despite the lack of any physical meaning for the $R^{1/4}$ model, some 
authors have actually restricted the inner and outer radial extent of 
elliptical galaxy light-profiles (e.g., Burkert 1993), 
or adjusted the sky-background level (e.g., Tonry et al.\ 1997), 
in order to make the $R^{1/4}$ model fit --- such is the ingrained 
belief in this classical model.  Other authors have chosen not to vary 
the S\'ersic exponent because they find $R_e$ and $\mu_e$ measurements 
are more stable if $n$ is fixed.  Unfortunately, this has nothing 
to do with ensuring the accurate recovery of the true half-light 
radius and surface brightness; setting $n=3$, for example, will have 
similar results on the perceived stability. 

Schombert (1986) recognized 
the inadequacy of the $R^{1/4}$ model to describe luminous Es, 
finding it only fitted the middle part $21 < \mu_B < 25$ of bright 
galaxy profiles.  Djorgovski \& Kormendy (1989) noted that it fits 
best elliptical galaxies with $M_B \sim -21$; brighter and fainter 
galaxies having different curvature than described by the $R^{1/4}$ 
model. 
That is, although real, luminosity-dependent, departures 
from the $R^{1/4}$ model have been known to exist for well 
over a decade, unfortunately, even today, they are often 
not even explored.  However, as we will see, such deviations 
provide the key to understanding the true nature/connection 
between dE and E galaxies. 

Studying a sample of 80 early-type galaxies in the Virgo and Fornax 
Clusters with the S\'ersic model, Caon, Capaccioli, \& D'Onofrio's (1993) 
and D'Onofrio, Capaccioli, \& Caon's (1994) exploration of how the elliptical 
galaxy profile shapes (n) vary systematically with the {\it model-independent} 
half-light galactic radii (and luminosity) has been monumental in advancing 
our understanding of elliptical galaxies --- placing it on a quantitative 
footing.  Furthermore, by including a sample of 187 dwarf 
galaxies, modelled by Davies et al.\ (1988) using this same generalized 
profile, Caon et al.\ (1993) showed that one continuous trend between 
galaxy size and profile shape spanned both galaxy classes.  Graham 
et al.\ (1996) subsequently showed this same trend to continue for even 
the brightest cluster galaxies.  
The shape of an elliptical galaxy's light-profile has since been 
shown to correlate strongly with other galactic quantities which have also 
been obtained independently of the S\'ersic model.  
For example, Graham et al.\ (2001b, 2002b) showed how 
the logarithm of $n$ correlated with a galaxy's central velocity 
dispersion (correlation coefficient $r=0.8$). 
Graham et al.\ (2001a, 2002a) showed the logarithm of $n$ to correlate 
($r=0.9$) with a galaxy's supermassive black hole mass.  
Different profile shapes are therefore obviously real, and can be 
parameterized with the S\'ersic $R^{1/n}$ model.  Indeed, just as de Vaucouleurs 
\& Capaccioli (1979) showed that beyond 10$\arcsec$, $n=4$ provided a good fit to 
the profile of NGC~3379 (spanning 10 mag of surface brightness), 
Bertin, Ciotti, \& Del Principe (2002) have shown that very different 
values of $n$ provide equally good fits to the extended profiles of 
other elliptical galaxies. 
Far from a dichotomy, at least in terms of galaxy profile shape and 
luminosity, dwarf and bright elliptical galaxies are extensions of each 
other (Cellone, Forte, \& Geisler 1994; Young \& Currie 1994, 1995, 1998; 
Durrell 1997; Gerbal et al.\ 1997; Graham \& Colless 1997; Jerjen 
\& Binggeli 1997; Prugniel \& Simien 1997; Trujillo, Graham, \& Caon 2001). 


In Section 2 we introduce our sample of 18 Coma Cluster, dE galaxy 
candidates imaged with the Hubble Space Telescope (HST) WFPC2 camera.  
We describe the data reduction and profile extraction.  Possibly for 
the first time, we show that the S\'ersic model, combined with 
an optional central point-source or central Moffat-convolved Gaussian, 
provides a good description to the full radial extent of 
{\it HST}-resolved dE galaxy light-profiles.  The structural parameters 
from the fitted models will be used in an exploration of the Fundamental 
Plane for dE galaxies (Guzm\'an et al.\ 2003, in prep). 
In Section 3 we confirm the trend between profile shape and 
luminosity found by others who had previously excluded the nuclear 
region in their modelling.  
We also explain the relation between inner profile slope ($\gamma\prime$) 
and luminosity as a consequence of the relation between global profile 
shape $n$ and luminosity.  We additionally observe a relationship between 
the luminosity of the underlying galaxy and the luminosity of the central 
component in dE galaxies.  Finally, in Section 4, we show how the 
difference between $\mu_0$ and $<\mu>_e$ (and also $\mu_e$) varies with 
light-profile shape and consequently with galaxy luminosity. 
Combining our dE data with a larger sample of literature data, 
we subsequently reveal how a continuous and linear relation between 
$\mu_0$ and luminosity results in an apparent dE/E dichotomy when one 
plots either $<\mu>_e$ (or $\mu_e$) against luminosity.  An apparent 
dichotomy is thus expected in the $<\mu>_e$--$\log R_e$ diagram.  
A summary is given in Section 5. 
Throughout this paper we use $H_0=70$ km s$^{-1}$ Mpc$^{-1}$.

\section{Data}

\subsection{Galaxy Sample} 

There have been many recent studies of the surface brightness 
profiles of Coma cluster galaxies (e.g., Gerbal 1997; Mobasher et al.\ 1999; 
Khosroshahi, Wadadekar, \& Kembhavi 2000; Gavazzi et al.\ 2000; 
Komiyama et al.\ 2002; Trujillo et al.\ 2002). 
These studies used ground-based images.  As our interest 
lies in the properties of dE galaxies having 
half-light radii of only a couple of arcseconds, we have 
obtained archival {\it HST} images.  Moreover, dE galaxies 
are often nucleated; failing to separate the excess nuclear light 
from the underlying galaxy light can severely bias the structural 
parameter analysis (e.g., Balcells et al.\ 2003). 
Although the Coma cluster is more distant than, for example, 
the Virgo and Fornax clusters, due to the high resolution of 
the {\it HST} we have been able to acquire surface brightness 
profiles of a quality better than typical ground-based profiles 
of Virgo and Fornax dE galaxies. 
Additionally, one of us (R.G.) has already studied the structural, 
chemical, and dynamical properties of Coma's brightest cluster 
members (Guzm\'an, Lucey, \& Bower 1993). 

Our Coma cluster dE galaxy candidates were chosen using 
$U$-, $B$-, and $R$-band images of the central 
$20\arcmin\times20\arcmin$ region obtained with WIYN/MiniMo and the 
INT/Wide Field Camera.  A detailed discussion of the selection
criteria is provided in Matkovi\'c \& Guzm\'an (2003, in prep).  
The selection criteria can be summarized as follows. 
Firstly, a color cut was applied in the $(U-B)$ vs.\ 
$(B-R)$ diagram such that $0.2 < (U-B) < 0.6$ and $1.3 < (B-R) <
1.5$.  Galaxies in this sample follow the same $(U-B)$ and $(B-R)$
color-magnitude relations defined by the bright ellipticals in Coma. 
According to galaxy evolutionary synthesis models, this color cut 
also minimizes contamination from background field disk galaxies at
z$\sim$0.2, which may affect galaxy samples at faint magnitudes.
Secondly, a limiting magnitude cut was applied such that $17.5 <
m_B < 20.5$.  This was applied in order to select ``dwarf'' galaxies 
with $-17.5 < M_B < -14.5$ in the Coma cluster --- 
assuming Coma to be at a distance of 100 Mpc. 
Spectra for a sub-sample of 
these Coma cluster dE-candidates were obtained using WIYN/HYDRA.  The
18 galaxies presented in this paper correspond to all the objects with
{\it HST}/WFPC2 F606W images in this sub-sample whose redshifts confirmed
their Coma cluster membership.  Despite being bonafide cluster members,
we continue referring to them as Coma cluster dE candidates since some
of these objects turn out to have morphologies more consistent with
being dwarf disk galaxies, as discussed below.

Table 1 provides the identification of our galaxy sample, as denoted in
the Godwin, Metcalfe \& Peach's (1983) Coma galaxy catalog.  
Figure~\ref{fig2} presents a mosaic containing the 18 {\it HST} Archival 
F6060W images of our sample of dE galaxy candidates in the Coma Cluster.  
The images were obtained from the {\it HST} Archive. The two main original
programs from which these images were taken were aimed at studying the
central cores, and the globular cluster populations, of bright
elliptical galaxies in Coma.  The elliptical galaxies were centered in
the Planetary Camera.  Our dE galaxy candidates were found in the three
Wide Field chips, i.e., at an angular distance of less than $\sim$1.5$\arcmin$ 
from a bright elliptical galaxy.

\subsection{Data reduction and profile fitting}

The {\it HST} images were reduced following the standard {\it HST} 
pipeline, combined and cleaned of cosmic-rays using crreject, further 
cleaned of cosmic rays using \texttt{LACOS} (L.A.COSMIC, van Dokkum 2001), 
and background 
subtracted using the wavelet decomposition method by Vikhlinin et al.\ 
(1998).  This variable large-scale subtraction was needed in order 
to define an accurate asymptotic value for the curve of growth that is
otherwise not well-defined due to the contamination of nearby galaxy
halos for many of our objects.

Photometry was performed using the \textsc{iraf} task \texttt{ELLIPSE} 
with both the galaxy center (determined with the task \texttt{CENTER}) 
and the position angle (measured at $\sim$2.5$\arcsec$, roughly 1 
$R_e$) fixed.  The ellipticity was allowed to vary.  The position 
angle was, however, allowed to vary for a couple of galaxies (GMP~3625 
and GMP~3629; Godwin, Metcalfe, \& Peach 1983) 
where $\sim$60-80$^{\rm o}$ differences 
exist between the position angles of the inner and outer isophotes. 
GMP~3629 is almost perfectly round and so these changes are 
not surprising, GMP~3625 has an ellipticity of 0.17-0.40. 

Surface brightness profiles and curves-of-growth were derived with the 
key assumption that the curve-of-growth had to flatten at large radii, 
i.e., the asymptotic value is well-defined.  In some cases this 
required a small constant background subtraction to remove residual 
sky.  Only in those cases where the galaxies were so close to the edge 
of the chip that the profile could not be measured over the entire 
galaxy was the condition of a flat curve-of-growth relaxed. In these 
cases, the sky was estimated as the median value at large distances 
from the galaxy.  We note that an unambiguous estimate of the 
sky-background is not always possible and the results may be affected 
by a certain subjectivity in 
the way the background was subtracted.  However, results from 
independent surface brightness profiles obtained from different images
of the same galaxy are in good agreement with each other (see later). 

The surface brightness profiles are presented in Figures~\ref{fig3} 
and \ref{fig4}, shown as a function of the geometric mean radius 
$R=R_{\rm maj}\sqrt{1-\epsilon}$.  As we had multiple profiles for 
each galaxy, the most typical have been shown.  The error bars shown 
on the profile data represent the errors associated to photon-noise and 
read-out-noise.  Systematic errors in the sky-background associated 
with residual contamination by the halos of other galaxies are not included. 
We model individual profiles as a combination of 
a PSF-convolved $R^{1/n}$ bulge model (Trujillo et al.\ 2001) and either a point-source or 
a PSF-convolved central Gaussian.  An optional PSF-convolved outer 
exponential model was used for galaxies with an obvious two component 
structure (in addition to the nuclear source). 
The Moffat PSF, and hence the point source profile, was measured from 
nearby stars or globular clusters on each chip. 
For any given galaxy, all the structural parameters were obtained 
simultaneously.  
We have verified that every galaxy's 
major-axis profile required the same composition of components as the 
geometric mean profiles, confirming our final component selection 
shown in Figures~\ref{fig3} and \ref{fig4}.  
No signal-to-noise ({\it S/N}) weighting was employed as it can result 
in poor fits when the inner profile contains additional components 
which are not considered in the modelling process.  
For example, if we only fitted {\it S/N}-weighted $R^{1/n}$ models, the 
presence of the nuclear components would severely bias 
the $R^{1/n}$ fits, producing erroneously high values of $n$.  

We have estimated the errors on the structural parameters by 
modelling different images of the same galaxy.  
Sometimes the exposure times were different, other times the galaxy 
appeared on a different WFPC2 chip.  Our error estimates therefore 
include the effects of: shot-noise, placement on the chip, estimation of the 
galaxy orientation and center, and the associated influence of pixelation. 
Additionally, by modelling the sky-background in different ways 
we have been able to further gauge the likely accuracy of the fitted 
model parameters.  
In all cases when we had multiple light-profiles, which we did for 
most galaxies, 
the surface brightness profiles were consistent with each other 
down to a level of $\mu_{\rm F606W}\sim 25$ mag arcsec$^{-2}$. 
Fitting models down to this surface brightness level we found the 
differences in the best-fitting S\'ersic 
parameters obtained from different profiles of the same 
galaxy spanned $\pm$0.05 mag arcsec$^{-2}$ in $\mu_e$, 
$\pm$5\% in $R_e$, and $\pm$4\% in $n$.  
Most profiles are not consistent with the $n$=1 exponential model.

\subsection{Disks}

\begin{figure}
\epsscale{0.75}
\plotone{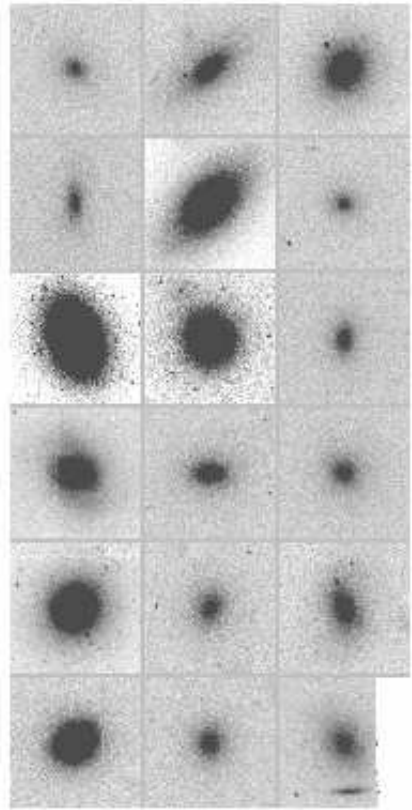}
\caption{Mosaic of the 18 dwarf elliptical galaxy candidates imaged 
using the F606W filter on the {\it HST} WFPC2.  
The galaxies are ordered by increasing GMP number 
(see Tables 1 and 2) 
along rows from the top left to the bottom right. 
Each sub-image is $12\arcsec\times12\arcsec$.  
The apparent differences in the background sky level are due to 
differences in the exposure time.  
}
\label{fig2}
\end{figure}

\begin{figure}
\epsscale{0.75}
\plotone{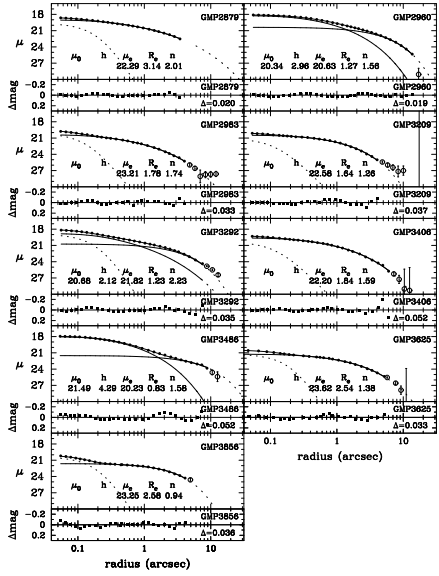}
\caption{
Geometric mean-axis ($R=\sqrt{ab}$) 
surface brightness profiles for some of the galaxy sample
listed in Table 1 and 2.  Every profile has been fitted with a Moffat-convolved
S\'ersic model (solid line).  Three profiles are additionally fitted 
with an outer exponential (also a solid line).  
An inner point-source, when detected, is shown by a dotted line.  
The outer extrapolated model is also shown by a dotted line.  
Only the filled circles were used in the modelling process, the 
larger open circles were not.  
The lower panel displays the residuals of the data about the fitted 
model.  The mean residual from the fit is given by $\Delta$ mag. 
}
\label{fig3}
\end{figure}

\begin{figure}
\epsscale{0.75}
\plotone{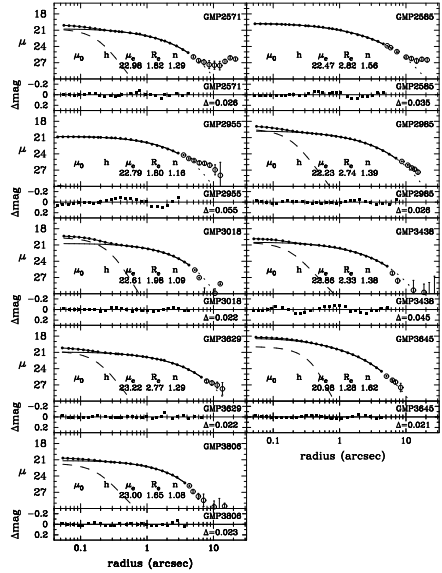}
\caption{
Similar to Figure~\ref{fig3} except that a 
Moffat-convolved Gaussian (dashed line) was used, instead 
of a point-source, to model the nuclear component in the 
remaining galaxy sample.
}
\label{fig4}
\end{figure}

Before proceeding, we checked for the presence of dwarf S0 galaxies 
(Sandage \& Binggeli 1984; Binggeli \& Cameron 1991) in our sample. 
From the image and surface brightness profile of GMP~3292 
it is clearly not a dE galaxy; there is evidence of faint outer 
spiral arms (Figure~\ref{fig2}), and it has a two-component 
light-profile which is not due to nucleation.  With a recessional 
velocity of 4955 km s$^{-1}$ (Matkovi\'c \& Guzm\'an 2003, in prep), 
this galaxy is not a background 
galaxy.\footnote{Redshifts 
have been obtained for our entire galaxy sample; all galaxies have 
velocities in the range from 4,000 to 10,000 km s$^{-1}$ and reside 
within 20 arcmin from the cluster center, consistent with being 
cluster members (Kent \& Gunn 1982).  The mean velocity and velocity 
dispersion of the Coma cluster are respectively $\sim$7,000 km s$^{-1}$ 
and $\sim$1,000 km s$^{-1}$ (Edwards et al.\ 
2002).} 
Its central surface brightness is brighter than $\mu_B=24$ mag 
arcsec$^{-2}$, and so technically it is not a dwarf spiral galaxy, as 
defined by Schombert et al.\ (1995).  
Due to the significant prominance of its bulge, it is also not 
a small late-type disk galaxy. 
GMP~3292 is thus likely to be a dwarf S0 galaxy. 
Unlike the extraordinarily faint inner spiral arms found by 
Jerjen, Kalnajs, \& Binggeli (2000) in the (previously 
classified) dE,N galaxy IC~3328, the arms in GMP~3292 are actually 
quite visible in the outer 
parts of GMP~3292.  An examination of the type performed
by Jerjen, Kalnajs, \& Binggeli (2000b) and Barazza, Binggeli, 
\& Jerjen (2002) is beyond the intended scope of this paper 
and will be left for future investigation.  Our sole objective 
here is to identify and exclude possible two component galaxies 
(not counting nucleation as a component). 

In addition to GMP~3292, we will exclude two further galaxies.  
Neither GMP~2960 nor GMP~3486 can be described with a global 
S\'ersic model plus some small nuclear component.  
The {\it FWHM} of the best-fitting central Gaussian, 
when simultaneously fitted with a global S\'ersic model, 
is greater than 0.5$\arcsec$, or $\sim$235 pc, for both of 
these galaxies.  Figure~\ref{fig3} shows both of these profiles 
fitted with an inner S\'ersic model plus an outer exponential
model.  We are unable to say whether the outer exponential is that
of a flattened disk, as in the case of a dwarf S0 galaxy,  
or whether the outer exponential is actually the underlying host galaxy 
and the inner S\'ersic model is describing an excessively large 
nuclear star cluster.  The latter scenario might represent the 
evolved state of blue compact dwarf galaxies (e.g., Papaderos 
et al.\ 1996; Doublier, Caulet, \& Comte 1999; Cair\'os et al.\ 2001, 
and references therein), although this suggestion is purely 
speculative on our part.  We do note that the ellipticity in GMP~2960 
changes from 0.06 at 1.27 arcseconds (the half-light radius of the inner
S\'ersic component) to 0.51 at 13.3 arcseconds (the outer most 
data point modelled by us), suggestive, albeit not conclusive, of 
an inclined outer disk.  For GMP~3486, the ellipticity decreases 
with radius (see Table 2). 

Lastly, GMP~3629 may also possess an outer disk; 
there is the suggestion of faint spiral 
arms in Figure~\ref{fig2}.  However, the width of the nuclear component 
in GMP~3629 is only 0.12$\arcsec$ ($\sim$ 60 pc), 
possibly too small to be considered a bulge.  We do however note the 
the alternative possibility 
that the disk may not actually dominate until radii greater than 
10$\arcsec$, the inner profile (0.1--10$\arcsec$) dominated by the 
bulge light and the central 0.1$\arcsec$ dominated by a nuclear star 
cluster.  
We feel, however, to conclude there is a disk beyond 
10$\arcsec$ would place too much faith in the outer, 
low surface brightness profile.  
%
Nonetheless, if this is the case, we found that the S\'ersic parameters 
from such a bulge/disk decomposition are very similar to those already 
given in Table 2. 
Similarly for GMP~2955 and a couple of other galaxies, 
a diffuse outer envelope/disk/halo of stars may exist beyond our 
surface brightness cutoff at $\mu_{F606W}\sim 25$ mag arcsec$^{-2}$. 
However, including this (potential) outer exponential component 
in the modelling process did not substantially modify the parameters 
of the more luminous inner S\'ersic component.  

Table 2 provides the structural parameters and some derived quantities 
for our sample.  No surface brightness corrections have been applied 
here.  A quick and dirty comparison of our total (extrapolated to 
infinity) $F606W$ magnitudes versus the photographic $b$-band 
magnitudes (within the $\mu_b=26.5$ mag isophote) from Godwin 
et al.\ (1983) is shown in Figure~\ref{figX}.  The agreement is 
encouraging, especially given that the color term is likely 
to be different for each galaxy.  The line which is shown 
simply represents a constant color difference of 1.0 mag.  
The potential disk galaxy GMP~3486 is the only notable outlier, 
some 0.5 mag deviant, and (along with GMP~3292 and GMP~2960) 
it will be excluded from here on.  This leaves us with a sample 
of 15 dE galaxies.  

In the figures which follow, we have corrected for $(1+z)^4$ 
dimming ($-$0.10 mag), $K$-correction (0.02 mag; Poggianti 1997), 
and Galactic extinction ($-$0.02 mag; Schlegel, Finkbeiner, \& 
Davis 1998, via NED\footnote{NASA Extragalactic Database.}). 

The Nuker model (Lauer et al.\ 1995) was not used because it can only 
describe a restricted portion of the inner light-profile.  Moreover, 
as first 
suggested in Graham et al.\ (2002b), Trujillo et al.\ (2003, in prep) 
reveal that the low-luminosity (i.e., the so-called ``power-law'') 
elliptical galaxies are in fact actually pure S\'ersic-law galaxies.  
That is, excluding for the moment the presence of additional nuclear 
components, the 3 S\'ersic parameters describe the entire light-profile
of these galaxies.

\section{Correlations amongst the dE structural parameters}

\begin{figure}
\epsscale{0.75}
\plotone{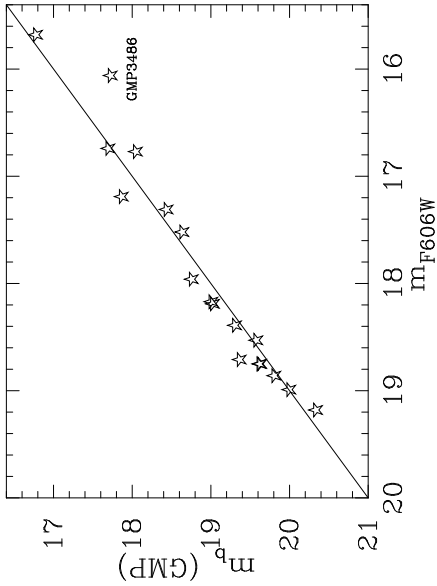}
\caption{
Model-independent, photographic $b$-band apparent magnitudes 
within the isophote $\mu_b=26.5$ from Godwin et al.\ 1983 (GMP) 
are plotted against our S\'ersic-model-dependent total $F606W$ 
galaxy magnitudes. The scatter is due, in part, to varying color 
differences for each galaxy.  The line drawn assumes a 
constant color difference of 1.0 mag.  The three brightest 
galaxies (GMP~2960, 3292, 3486) are potential disk galaxies.  
}
\label{figX}
\end{figure}

\begin{figure}
\epsscale{0.75}
\plotone{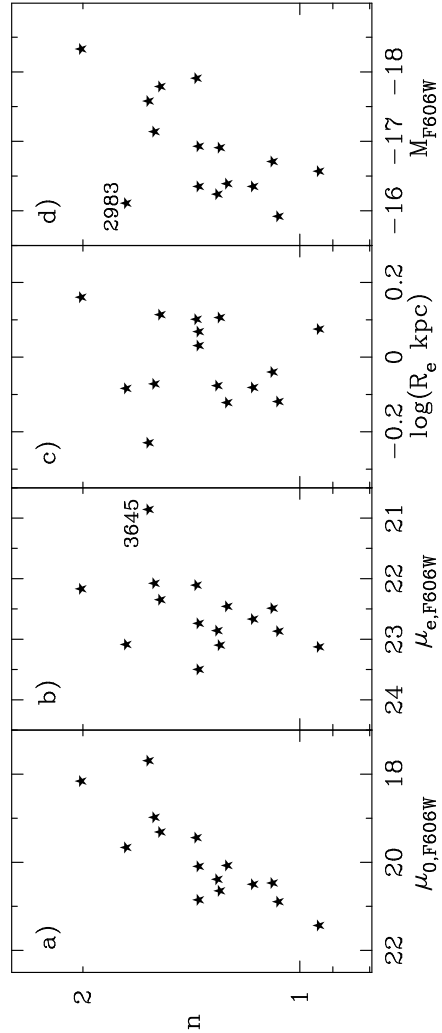}
\caption{
S\'ersic indices $n$ are shown on a logarithmic axis 
against: (a) the central surface brightness ($\mu_0$) of the underlying 
host galaxy; 
(b) the surface brightness ($\mu_e$) at the effective radius of the 
host galaxy $R_e$; 
(c) the effective radius $R_e$; and (d) the absolute magnitude of the 
host galaxy as derived from the fitted S\'ersic model. 
Surface brightnesses and magnitudes are those obtained with 
the F606W filter, as are the values for $R_e$ and $n$ (see Table 2).  
Typical errors on the value of $n$ and $R_e$ are $\pm$4\% and $\pm$5\%
respectively.  Typical errors for the values of $\mu_0$ and $\mu_e$ are 
$\pm$0.05 mag. 
}
\label{fig5}
\end{figure}

The present dE galaxy sample spans only 2.4 mag in absolute 
magnitude (excluding the brightest 3 galaxies in Figure~\ref{figX}, 
which are possible disk galaxies).  
Consequently, intrinsic scatter may dominate
many parameter correlations.  Nonetheless, in order to inspect 
whether the current data set displays the same general trends 
as other dE galaxies, Figure~\ref{fig5} presents the logarithm 
of the S\'ersic index $n$ plotted against the host galaxy's: 
(a) central surface brightness $\mu_0$; 
(b) effective surface brightness $\mu_e$;  
(c) logarithm of the effective radius $R_e$; and
(d) absolute magnitude in {\it F606W}.  
By the term `host galaxy' we mean the underlying galaxy --- 
free of the flux from additional nuclear components (which are
typically $\sim$1\% of the host galaxy flux).  
All quantities have been 
derived from the S\'ersic fit, the quality of which can be 
judged in Figures~\ref{fig3} and \ref{fig4}. 
{\it Importantly, the 
agreement between our model-dependent magnitudes and the 
model-independent values from Godwin et al.\ (1983) 
(see Figure~\ref{figX}), reveals that the correlation 
between profile shape ($n$) and magnitude in Figure~\ref{fig5}d 
is not an artifact from our use of S\'ersic models.}  
However, 
our limited range in luminosity and small sample size appears 
inadequate to reveal any strong trend between $\log n$ 
and $\log R_e$ (e.g., Caon et al.\ 1993; Young \& Currie 1995).  
The most notable correlation is between $\log n$ and $\mu_0$; 
the Pearson correlation coefficient is -0.83 
at a significance of 99.99\%.  
Modelling the bulges of spiral galaxies, other authors have found 
a similarly strong trend (e.g., Khosroshahi et al.\ 2000; 
M\"ollenhoff \& Heidt 2001).  
A reduced-major-axis regression analysis (Feigelson \& Babu 1992) gives
$\mu_{\rm 0,F606W} = 21.49(\pm0.28) -11.90(\pm2.33)\log(n)$, 
where the uncertainties have come from a jackknife sampling of the data. 
Excluding GMP~2983, 
the Pearson correlation coefficient between $\log n$ 
and absolute host galaxy magnitude is $r=-0.77$, at a significance of 99.9\%. 
An orthogonal regression analysis gives
$M_{\rm gal,F606W} = -15.88(\pm0.29) -8.45(\pm1.60)\log(n).$ 

The sense of these trends agree with the correlations 
shown by Cellone et al.\ (1994), Young \& Currie (1994), Jerjen \& 
Binggeli (1997), Durrell (1997), Ryden et al.\ (1999), Schwarzkopf \& 
Dettmar (1997), and others. 

\subsection{Nuclei}

Many dE galaxies are known to be nucleated (e.g., Sandage \& Binggeli 1984; 
Binggeli et al.\ 1984; Ferguson \& Sandage 1989).  
However, most {\it HST} studies of nearby, nucleated galaxies have focused 
on spiral galaxies (e.g., Phillips et al.\ 1996; Carollo et al.\ 1997a; 
Matthews et al.\ 1999; B\"oker et al.\ 2002; Carollo et al.\ 2002; 
Balcells et al.\ 2003) or luminous elliptical galaxies (e.g., Lauer et al.\
1995; Ravindranath et al.\ 2001; Rest et al.\ 2001).  To the best of 
our knowledge, there has been no structural analysis of the nuclear
star clusters in dE galaxies using the {\it HST}. 

\begin{figure}
\epsscale{0.75}
\plotone{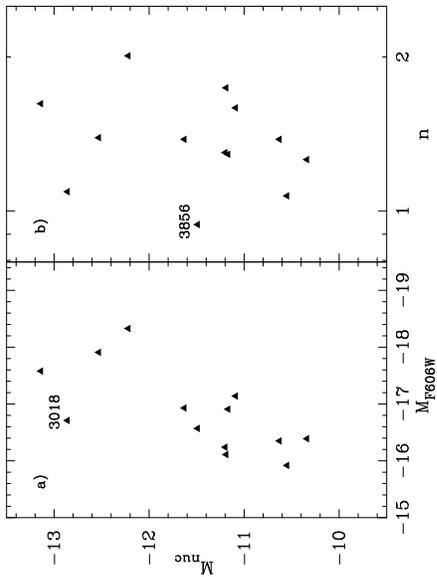}
\caption{The F606W magnitude of the central component ($M_{\rm nuc}$) 
versus a) the magnitude of the host galaxy, and b) the global 
profile shape $n$.}
\label{fig6}
\end{figure}

Figure~\ref{fig6} shows the magnitude of the nuclear component 
plotted against: 
(a) the absolute host galaxy magnitude; and 
(b) the logarithm of the galaxy profile shape $n$.  
It is noted that no central component was detected in GMP~2585 
and GMP~2955, and these galaxies are therefore not included in 
this figure. 
There is a fair amount of scatter in these sparsely populated 
diagrams.  Although the data sample is small, we can however 
remark that it does reflect the apparent trend seen between 
nuclear component magnitude and host galaxy magnitude shown 
in Figure 7 of Phillips et al.\ (1996; excluding the star-forming 
knots in that diagram).  There is, however, a difference in what 
these diagrams show, since those authors used the total 
{\it (bulge-plus-disk)} magnitudes from a sample of nearby 
disk galaxies (see also Carollo 
et al.\ 1998).  More in-line with what we show in Figure~\ref{fig6}, 
Balcells et al.\ (2003) have recently shown the magnitudes of 
the nuclear components in the bulges of early-type disk galaxies 
correlate strongly ($r_s$=0.77) with these galaxy's bulge magnitudes.

An orthogonal regression analysis on Figure~\ref{fig6}a yields 
$M_{\rm nuc}=(1.37\pm 0.55)(M_{\rm gal}+17.5) - (12.43\pm 0.55)$
using the F606W filter, 
which is equivalent to 
\begin{equation}
L_{\rm nuc} = 10^{4.97\pm 0.22}\left(\frac{L_{\rm gal}}{10^7}\right)^{1.37\pm 0.55}. 
\label{point}
\end{equation}
To explore the stability of this relation, and obtain 
a more robust result, we have re-derived this correlation 
after excluding two potential outlying galaxies; 
specifically, those with the brightest nuclear components 
(GMP~3018 and GMP~3645): 
\begin{equation}
L_{\rm nuc} = 10^{4.76\pm 0.10}\left(\frac{L_{\rm gal}}{10^7}\right)^{0.87\pm 0.26}.
\label{punt}
\end{equation} 
More data is of course required before 
Equations~\ref{point} or \ref{punt} can be considered 
universal for nucleated dE galaxies.  
Environment may also play a role.

\subsection{Inner profile slope} 

Having modeled the nuclear component, we can now explore the inner slope 
of the underlying host galaxy profile.  Interest in this quantity arose 
from the discovery of a bimodal distribution of slopes for 
E galaxies more luminous that $M_B\sim -18$ (Ferrarese et al.\ 1994; Lauer et 
al.\ 1995; Gebhardt et al.\ 1996), and the implication of a different galaxy 
formation process.  

For a S\'ersic model, the negative logarithmic slope of the 
light-profile at any radius $R$ can be given by the expression 
\begin{equation}
\gamma\prime(R) \equiv -d\log I(R)/d\log R =\frac{b}{n}\left(\frac{R}{R_e}\right)^{1/n}. 
\end{equation}  
For a fixed $R/R_e$ ratio, $\gamma\prime$ is thus a monotonically increasing 
function of $n$ (Graham et al.\ 2002b; their Figure 4). 
{\it Given the correlation between $n$ and magnitude (Figure~\ref{fig5}d), 
$\gamma\prime$ is therefore obviously correlated with magnitude.} 

\begin{figure}
\epsscale{0.75}
\plotone{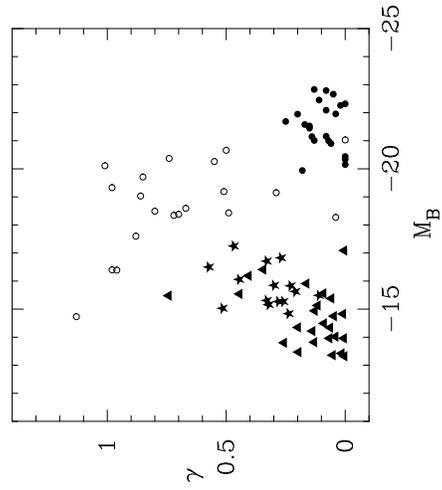}
\caption{
The negative logarithmic slopes ($\gamma$) of the underlying dE host 
galaxy light-profiles (measured at $R=0.2\arcsec$) versus the 
host galaxy magnitudes are shown as stars.  A ($B-F606W$) color of 
1.08 has been used.  
The dE galaxy data from Stiavelli et al.\ (2001) are shown as triangles; 
their value of $\gamma$ is a measure of the average (fitted) Nuker model 
slope between 0.1 and 0.5$\arcsec$.  A ($B-V$) color of 0.9 was assumed 
by us to convert their magnitudes to the $B$-band.  
The filled and open circles are respectively the `core' and `power-law' 
elliptical galaxies from Faber et al.\ (1997).  Excluding the two bottom 
middle outliers, one can clearly see that $\gamma$ increases as galaxy 
magnitude brightens, until the detection of partially evacuated cores 
in the most luminous ($M_B\lesssim 20.5$ mag) ellipticals. 
}
\label{fig7}
\end{figure}

Rather than use $\gamma\prime(R/R_e=constant)$, that is, 
ignoring differences in galaxy size, 
Figure~\ref{fig7} shows a plot of $\gamma\prime(R=0.2\arcsec)$ 
versus the host galaxy magnitude.  
Although we recognize this is perhaps not the best estimate of 
the inner profile slope, as it measures the slope at different 
fractions of the half-light radius, and identical galaxies in 
clusters at different distances will be sampled at different 
physical radii, it is nonetheless somewhat comparative with 
past investigations. 
We have used a $B-${\it F606W} color of 1.08 (Fukugita, Shimasaku, 
\& Ichikawa 1994) in order to present $B$-band magnitudes for our galaxy 
sample.  
The profile slope $\gamma\prime$ was derived from the fitted S\'ersic 
model and therefore explicitly avoids any potential bias in the 
slope from additional nuclear components.  To figure~\ref{fig7} we 
have added the dE galaxy data from Stiavelli et al.\ (2001).  
Following Carollo et al.\ (1997b), they computed the 
average logarithmic profile slope from 0.1-0.5 arcseconds. 
Conversion from the {\it HST} F555W filter used by Stiavelli 
et al.\ (2001) to the $B$-band used by the Nuker team 
was performed by us assuming a constant $B-F555W$ color of 0.9. 
The Nuker team's bright elliptical galaxy sample (Faber et 
al.\ 1997), excluding their S0 galaxies, are also included here. 
In spite of the different methods and radial ranges used to 
obtain the inner profile slope, and inaccuracies from our 
assumption of constant color terms, 
there is a general correlation in Figure~\ref{fig7}
such that the inner profile slope 
steepens as the absolute magnitude of the host galaxy brightens. 
This trend, albeit with the odd outlier, can be seen to continue until 
the onset of core formation in the brightest elliptical galaxies.  

The observed cores in the luminous elliptical galaxies are thought to 
have arisen from the partial evacuation of the nuclear region by 
coalescing blackholes (e.g., Ebisuzaki, Makino, \& Okumura 1991; 
Makino \& Ebisuzaki 1996; Komossa et al.\ 2003).  Whatever process(es) 
have reduced these galaxies central surface brightness profiles,  
the high luminosity ``core'' galaxies clearly depart from the relation
defined by galaxies without cores. 
Presumably the depletion of the core occurs rapidly, hence the observed gap 
$0.3<\gamma\prime<0.5$ at the bright end of the relation (Ferrarese 
et al.\ 1994; Lauer et al.\ 1995; Gebhardt et al.\ 1996; Faber et al.\ 
1997), although it should be noted that some galaxies have since 
been found to reside here (Rest et al.\ 2001;  Ravindranath et al.\ 2001).

\section{Structural connections between dE and E galaxies}

To illustrate the continuity between the dE and E galaxies, and 
at the same time explain their apparently different behavior in 
certain structural parameter diagrams, 
we will use the large compilation of dE and E galaxies presented 
in Graham et al.\ (2002b, their Figure 1).  This collection of 232 
galaxies spans 10 mag in absolute magnitude and consists of the 
luminous E galaxies from the sample of Faber et al.\ (1997), the 
intermediate-to-bright E galaxies from Caon 
et al.\ (1993) and D'Onofrio et al.\ (1994), and the dE galaxy 
sample of Binggeli \& Jerjen (1998) and Stiavelli et al.\ (2001).  
Known lenticular galaxies are excluded.  One difference 
between the central surface brightness data shown in Graham et 
al.\ (2002b) and those shown here 
is that rather than use the observed (seeing-reduced) central 
surface brightness values from the sample of Caon et al.\ (1993) and 
D'Onofrio et al.\ (1994), we will use the best-fitting S\'ersic model
value.  However, because of core formation, we have excluded 
those galaxies from these Authors with $M_B<-20.5$ mag in the 
plots using central surface brightness. 
Finally, we have added our own sample of Coma dE galaxies.  
We additionally show correlations using $\mu_e$, $<\mu>_e$, 
$\log R_e$, and $\log n$, which are not shown in Graham et al.\ (2002b). 

The only {\it HST} study of dE galaxy surface brightness 
profiles that we are aware of is that by Stiavelli et al.\ (2001). 
After excluding the inner arcsecond of the profiles because of 
nucleation, they fitted exponential, $R^{1/4}$, and S\'ersic 
models to a sample of 23 Virgo and Fornax Cluster, and 2 Leo Group,  
dE galaxies\footnote{Stiavelli et al.\ (2001) additionally 
fitted Nuker models to the inner profile after excluding the 
innermost 0.5$\arcsec$.}.  From their error analysis, only 7 of 
these galaxies have profiles consistent (at the 2$\sigma$ 
level) with an exponential $n=1$ model (see their Table 2).

For the full galaxy sample, Figure~\ref{fig8} shows, when available, 
the absolute $B$-band galaxy magnitude plotted against three measures 
of surface brightness: the mean surface brightness $<\mu>_e$ 
within the effective half-light radius $R_e$; the surface brightness 
$\mu_e$ at the half-light radius; and the central surface 
brightness $\mu_0$ of the host galaxy (excluding the flux from nuclear 
components).  The middle panel shows these three surface brightness 
values versus the logarithm of the S\'ersic index $n$, while the lower 
panel shows these three values against the logarithm of the half-light 
radius $R_e$.  The value of $R_e$, and the absolute magnitudes, were 
obtained from the above papers and re-derived assuming a Hubble 
constant of 70 km s$^{-1}$ and a Virgo and Fornax distance modulus 
of 31.2. 

One of the most quoted papers to support the dE/E 
dichotomy is that of Kormendy (1985).  In that paper the author wrote that, 
``the most surprising result of this paper is that there is a 
large discontinuity between the parameter correlations for 
elliptical and dwarf spheroidals'', and concluded that ``dwarf
elliptical galaxies are very different from the sequence of 
giant ellipticals.''  
This conclusion was based largely on Figure 3 (top right panel) 
from that paper, 
which included 11 dwarf elliptical galaxies and a number of luminous 
elliptical galaxies.   Indeed, this plot of central surface 
brightness versus galaxy magnitude does show a large 
discontinuity between the dE and E galaxies.  
However, there is an absence of galaxies with 
magnitudes around $M_B=-18\pm1$ in this sample, exactly where
one might expect to see the two groups connect.  
Nearly a decade later, the Astronomy and Astrophysics Review paper  
by Ferguson \& Binggeli (1994; their Figure 3) had a big question 
mark as to ``how'' and indeed ``if'' the dE and E galaxies might 
actually connect in this diagram.  That is to say, the idea of a 
discontinuity has been around (and largely accepted) for many years. 

If we are to remove galaxies
having $M_B=-18\pm1$ from our diagram of $\mu_0$ vs.\ $M_B$ 
(Figure~\ref{fig8}c) we will obtain a figure which looks very 
much the same as Kormendy's figure.  {\it Thus, the answer 
to the believed dichotomy lies, in part, in the sample 
selection used by Kormendy (1985).  This explains the discontinuity 
but not the change in the slope. The answer to the latter resides in 
the observation that luminous elliptical galaxies possess partially 
evacuated ``cores''}.  Their central surface brightnesses are thought 
to be a modification of their original, more luminous value.  
As stressed in Graham et al.\ (2002b), the brighter galaxies lying 
perpendicular to the relation defined by the less luminous 
($M_B\gtrsim -20.5$) 
elliptical galaxies in the $\mu_0$-$M_B$ diagram are all ``core'' 
galaxies.  There is no dichotomy at $M_B\sim -18$ in this diagram.  
However, there is a ``dichotomy'' at $M_B\sim -20.5$; but this is 
not necessarily the result of a different initial formation 
mechanism for galaxies brighter or fainter than this value.  
Rather, (subsequent) core formation has apparently modified the 
central surface brightness in galaxies brighter than 
$M_B\sim -20.5$.  The initial mechanism(s) of dE and E galaxy 
formation are therefore likely to be the same given the 
continuity at $M_B=-18$.  
Importantly, luminous elliptical galaxies should not be viewed 
as the norm, but instead are the exception to the relation between 
central surface brightness and magnitude which exists over 
(at least) some 8 mag.  Indeed, analysing over 100 dwarf spheroidals
($M_B\gtrsim -13$ mag, Grebel 2001), Hilker, Mieske, \& Infante (2003) 
and Grebel, Gallagher, \& Harbeck (2003) show that this trend 
continues down to at least $M_B\sim -8$ mag.

A second reason why people have thought that luminous E galaxies 
are different from low-luminosity (dwarf) elliptical 
galaxies is because they don't follow the same $M_B$--$\mu_e$ 
(and $M_B$--$<\mu>_e$) relation, as seen in Figure~\ref{fig8}b 
(and Figure~\ref{fig8}a).  
Given that there is a continuous and linear relationship 
between $M_B$ and $\mu_0$, suggestive of a 
similar formation mechanism, until the subsequent(?) process of 
core formation, why is the relationship between $M_B$ and 
$\mu_e$ (and $<\mu>_e$) apparently different for the dE and E 
galaxies?  It turns out the reason for this has nothing to do with 
core formation but is due to the systematic changes in profile shape 
with galaxy magnitude (Figure~\ref{fig9}).  

For the S\'ersic model, 
\begin{equation}
\mu_e = \mu_0 + 1.086b
\end{equation}
and 
\begin{equation}
<\mu>_e = \mu_e -2.5\log[e^{b}n\Gamma(2n)/b^{2n}] 
\end{equation}
(see, e.g., Caon et al.\ 1994; Graham \& Colless 1997).  
Figure~\ref{fig10} shows the differences 
between $\mu_0$ and $\mu_e$, and $\mu_0$ and $<\mu>_e$, as a function 
of profile shape $n$. 

The line in 
Figure~\ref{fig9} is now used to determine a representative value 
of $n$ for a given $M_B$.  Using this correlation, the line shown 
in Figure~\ref{fig8}c can then be transformed into a relationship 
between $M_B$ and $\mu_e$ 
(and $M_B$ and $<\mu>_e$) and is shown in Figure~\ref{fig11}.  
One can immediately see that we have reproduced the observed 
correlations in Figure~\ref{fig8}a and \ref{fig8}b.    The 
apparently different slope for the dE and E galaxies in these 
diagrams is merely a consequence of a continuously varying profile 
shape with galaxy luminosity --- it certainly does not 
imply a different galaxy formation mechanism.   
Without this understanding, these diagrams had been a considerable 
red herring to our understanding of dE/E galaxy formation. 

We have discussed the behavior of the various luminosity -- 
surface brightness diagrams.  We will now quickly explain why 
these lead to a different behavior  for the dE and E galaxies in 
the $<\mu>_e$--$\log R_e$ plane.  In this diagram, a break, or 
change in slope, at $M_B\sim -20$ is known to exist 
(Kodaira, Okamura, \& Watanabe 1983; Binggeli et al.\ 1984; 
Binggeli \& Cameron 1991; Capaccioli et al.\ 1993). 
Modelling elliptical galaxies from the bright end of the 
luminosity function results in the Kormendy (1977) relation 
(e.g., Hoessel, Oegerle, \& Schneider 1987; Graham 1996), 
while modelling fainter elliptical galaxies is known to produce 
a different relation (e.g., Binggeli \& Cameron 1991).  
The different slopes have been interpreted in the past as 
evidence for a dE/E galaxy dichotomy.  
However, the difference observed in this diagram need not imply 
a different formation mechanism. 
From the relation $L=2\pi R_e^2<I>_e$, the diagram 
of $M_B$ versus $<\mu>_e$ (Figure~\ref{fig8}a) can be used to 
determine $R_e$.  One 
can then immediately understand why diagrams of $\log R_e$ versus 
$<\mu>_e$ (Figure~\ref{fig8}g) show a different trend for the dE 
and E galaxies.  The same explanation applies to the $M-\log R_e$ 
relation. 

In conclusion, there is a continuous structural relation between 
the alleged dE and E galaxy classes until core formation is detected. 
The use of $\mu_e$ or $<\mu>_e$, instead of $\mu_0$, and the use of 
$R^{1/4}$ models, has blinded our realization of this for a long time.

\section{Summary}

This paper provides an analysis of {\it HST}-resolved dwarf 
elliptical galaxies modelled with a S\'ersic function 
and either a central point source or a resolved central Gaussian 
component.  That is, we 
have taken full advantage of the {\it HST} resolution and modelled 
the complete galaxy profiles.  We find that the structure of 
the dE galaxies is consistent with that found from studies which 
avoided the inner arcsecond (or three).  Moreover, the 
3-parameter S\'ersic model remains a good description of the 
{\it entire} (underlying) host galaxy light-profiles. 

Three of our initial 18 dE galaxy candidates may contain outer disks, 
or in any case are two component systems (in addition to possible 
nucleation); they are all possibly dwarf S0 galaxies. 
Thirteen of the remaining 15 dE galaxies are nucleated.  
Excluding two possible outliers, we found 
$L_{\rm nuc} = 10^{4.76\pm 0.10}\left( L_{\rm gal}/10^7\right) ^{0.87\pm 0.26}$. 
However, our galaxy sample spans only a little over 2 mag in 
absolute magnitude and hence an increased sample with a greater 
range of magnitudes would be valuable. 
Despite this, we find strong 
correlations between $\log n$ and $\mu_0$ ($r=-0.83$) and 
$\log n$ and $M_B$ ($r=-0.77$).  
The correlation between the inner, logarithmic profile slope
$\gamma\prime$ and $M_B$ is explained here as a consequence of the 
relations between $\gamma\prime$ and $\log n$, and $\log n$ and $M_B$. 
Galaxies fainter than $M_B\sim -20.5$ have progressively shallower 
inner profile slopes. 

By including 232 E galaxies from the literature, which (starting at 
$M_B\sim -13$) span 10 mag in absolute magnitude, we have shown that 
more luminous elliptical galaxies have 
brighter central surface brightnesses --- until the detection 
of core formation at $M_B\sim -20.5$. 
The linear relationships between magnitude, the logarithm of the 
profile shape $n$, and the central surface brightness, 
are used to predict and explain why the low- and high-luminosity 
elliptical galaxies display a different behavior in diagrams 
of luminosity versus effective and mean surface brightness.  
It has nothing to do with core formation, nor does it imply a 
different formation mechanism for each galaxy class, as has 
been the interpretation in the past.  It is instead a natural 
consequence of the previously mentioned linear relations. 
Contrary to popular belief, dE galaxies appear to be the 
low-luminosity extension of brighter E galaxies; there is no 
physical boundary at $M_B=-18$.  The smooth and continuous 
change in these galaxy's structural properties suggests that a 
similar physical process, or processes, have governed the 
evolution of the entire dE + E galaxy family --- with the notable
departure of only the most luminous ellipticals from the $M_B$--$\mu_0$ 
diagram at $M_B\sim -20.5$.   With this exception in mind, the 
mechanism of how dE and E galaxies collapsed to form stars is 
therefore expected to be similar. 

\begin{figure}
\epsscale{0.75}
\plotone{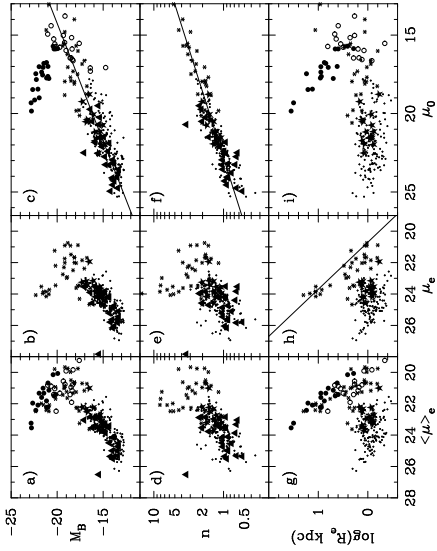}
\caption{The mean surface brightness within $R_e$ ($<\mu>_e$), the surface 
brightness at $R_e$ ($\mu_e$), and the central host galaxy surface brightness
($\mu_0$) are shown against the host galaxy magnitude ($M_B$), the global 
profile shape ($n$), and the half-light radius ($R_e$).  
Color terms are explained in the text.  
Due to biasing from the magnitude cutoff at $M_B\sim -13$, 
the line $M_B = (2/3)\mu_0 - 29.5$ in panel c) has been estimated
by eye rather than using a linear regression routine.  
The line $\mu_0=22.8-14\log(n)$ in panel f) has also been estimated by eye.  
The line in panel h) has a slope of 3 and represents the Kormendy (1977) 
relation known to fit the luminous elliptical 
galaxies which define the panhandle of this complex distribution 
(Capaccioli \& Caon 1991; La Barbera et al.\ 2002).  
Dots represent dE galaxies from Binggeli \& Jerjen (1998), 
triangles represent dE galaxies from Stiavelli et al.\ (2001), large stars 
represent our Coma dE galaxies, asterix represent intermediate to bright E 
galaxies from Caon et al.\ (1993) and D'Onofrio et al.\ (1994), open circles 
represent the so-called ``power-law'' E galaxies from Faber et al.\ (1997), 
and the filled circles represent the ``core'' E galaxies from these same 
Authors. 
The fundamental relations are between $\mu_0$, $\log n$, and magnitude; 
with an obvious modification of $\mu_0$ once a core forms in the 
brightest E galaxies.  
}
\label{fig8}
\end{figure}

\begin{figure}
\epsscale{0.75}
\plotone{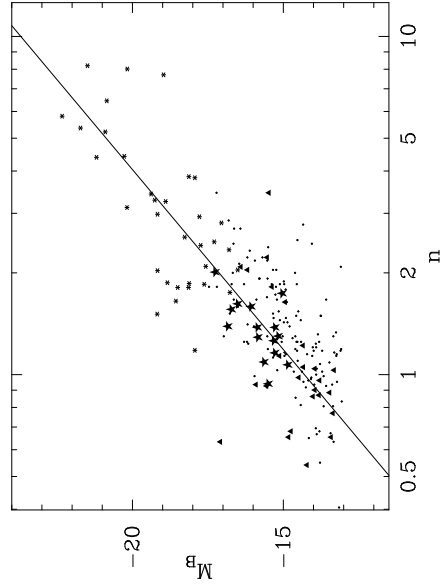}
\caption{Absolute $B$-band galaxy magnitude versus the logarithm 
of the S\'ersic shape index $n$.  All the galaxies shown in 
Figure~\ref{fig8}, except for the the Nuker team galaxies (Faber et al.\ 
1997) for 
which values of $n$ are not available, have been included.  
Due to biasing from the magnitude cutoff at $M_B\sim -13$, 
the line $M_B = -9.4\log(n) - 14.3$ has been estimated by eye.  
The central surface brightness values obtained from S\'ersic models 
fitted to luminous ($M_B\lesssim -20.5$) E galaxies follow this same 
relation (Jerjen, Binggeli, \& Freeman 2000), 
although they were not used to define it. 
}
\label{fig9}
\end{figure}

\begin{figure}
\epsscale{0.75}
\plotone{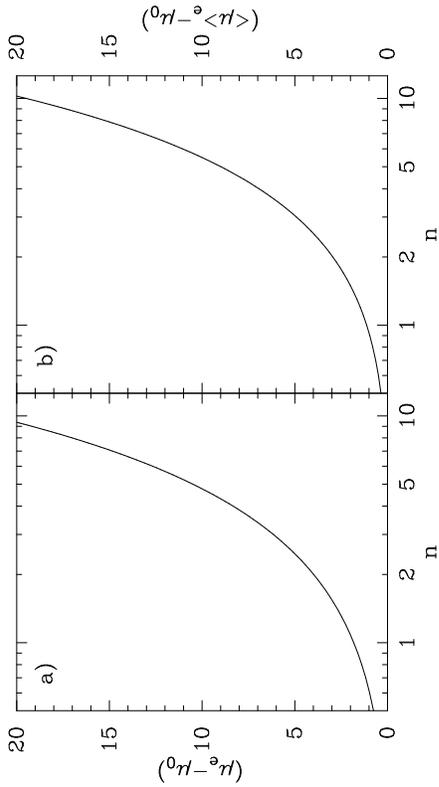}
\caption{Difference between the central surface brightness ($\mu_0$) 
and (a) the effective surface brightness ($\mu_e$), and (b) the 
mean surface brightness ($<\mu>_e$), 
as a function of profile shape $n$.}
\label{fig10}
\end{figure}

\begin{figure}
\epsscale{0.75}
\plotone{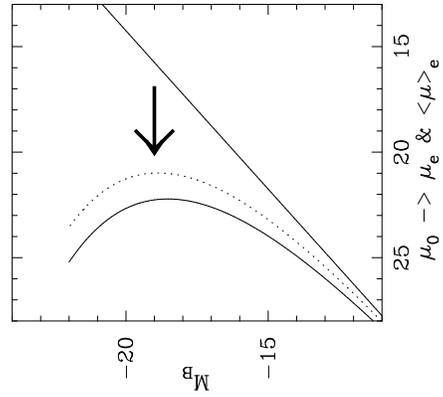}
\caption{Predicted change to the galaxy magnitude -- central surface 
brightness diagram when the mean ($<\mu>_e$, dotted curve) and 
effective ($\mu_e$, solid curve) surface brightness are used instead
of the central surface brightness ($\mu_0$, straight line). 
Derived from knowledge of how the profile shape $n$ varies with 
galaxy magnitude (Figure~\ref{fig9}) and how ($\mu_0 - <\mu>_e$) and 
($\mu_0 - \mu_e$) vary with $n$ (Figure~\ref{fig10}). 
Comparison with real data can be made by looking at 
Figure~\ref{fig8}a,b. 
}
\label{fig11}
\end{figure}

Simulations of elliptical galaxy catalogs which have used only 
$R^{1/4}$ models do not represent the real galaxy population
in our Universe.  Similarly, $N$-body merger models 
which finish with galaxies having a range of luminosities but with 
structural homology (e.g., all with $R^{1/4}$ profiles), have 
also not reproduced what is observed in nature.
It would be of great interest to re-examine whether the surface 
brightness profiles of elliptical galaxies simulated in an 
hierarchial clustering cosmology (e.g.\ Cole et al.\ 2000) 
display a range of profile shapes which vary with galaxy luminosity. 
Additionally, observational work which has {\it a priori} assumed 
all galaxies can be approximated with either an exponential surface 
brightness profile (e.g., Shao et al.\ 2003), or with 
an $R^{1/4}$ profile (e.g., Bernardi et al.\ 2003a,b)\footnote{The 
global curvature in the ($R^{1/4}$-model minus data) residual profiles 
in Bernardi et al.\ (2003a; their figure 7) clearly show deviations 
from the $R^{1/4}$-model which can be well matched using an $R^{1/n}$ 
model.}  
will certainly obtain a luminosity-biased set of surface brightness 
and galaxy size parameters which cannot reproduce the trends observed
in, for example, the top panel of Figure~\ref{fig8}. 
Lastly, 
models of bright elliptical galaxies built from the merging of 
elliptical galaxies fainter than $M_B\gtrsim-20.5$ must not assume 
the correlations defined by the brightest elliptical galaxies 
necessarily apply to the pre-merged galaxies.  
Unless merger models include the growth and merging of supermassive 
black holes, or some other mechanism of core depletion, they should 
predict brighter central surface brightnesses with increasing galaxy 
luminosity.

\acknowledgements
We wish to thank Ileana Vass for her help with the archival 
{\it HST} images. 
R.G.\ acknowledges funding from NASA grant AR-08750.02-A. 
This research has made use of the NASA/IPAC Extragalactic Database 
(NED) which is operated by the Jet Propulsion Laboratory,
California Institute of Technology, under contract with the 
National Aeronautics and Space Administration.

\clearpage

\clearpage
\begin{deluxetable}{lccccc}
\footnotesize
\tablecaption{Galaxy Sample. \label{tbl-1}}
\tablewidth{0pt}
\tablehead{
Gal. &  R.A. & Dec. &  Vel.        &  $m_b$ &  Type \\ 
     &       &      &  km s$^{-1}$ &  mag   &       \\
}
\startdata
GMP~2571 &13:00:37 &27:55:50.6  & 6103 & 19.81 & dE \\
GMP~2585 &13:00:35 &27:56:32.2  & 6898 & 18.44 & dE \\
GMP~2879 &13:00:11 &28:03:53.2  & 7271 & 18.05 & dE \\
GMP~2955 &13:00:06 &28:04:47.4  & 5069 & 19.62 & dE \\
GMP~2960 &13:00:05 &28:01:26.5  & 5847 & 16.78 & ?  \\
GMP~2983 &13:00:04 &28:00:28.8  & 6360 & 20.00 & dE \\
GMP~2985 &13:00:04 &27:57:51.3  & 5312 & 17.87 & dE \\
GMP~3018 &13:00:01 &27:59:27.8  & 7477 & 19.31 & dE \\ 
GMP~3209 &12:59:44 &28:00:45.1  & 7096 & 19.37 & dE \\
GMP~3292 &12:59:38 &28:00:01.8  & 4955 & 17.70 & dSp \\
GMP~3406 &12:59:30 &28:01:13.2  & 7114 & 18.76 & dE \\
GMP~3438 &12:59:29 &28:01:07.7  & 5942 & 19.01 & dE \\
GMP~3486 &12:59:25 &27:56:02.5  & 7522 & 17.73 & ?  \\
GMP~3625 &12:59:16 &27:53:07.7  & 6516 & 19.63 & dE \\ 
GMP~3629 &12:59:16 &27:53:55.2  & 5219 & 19.03 & dE \\
GMP~3645 &12:59:15 &27:53:42.4  & 6366 & 18.64 & dE \\
GMP~3806 &12:59:03 &27:58:27.4  & 5481 & 20.34 & dE \\
GMP~3856 &12:58:60 &27:59:34.5  & 6184 & 19.58 & dE \\ 
\enddata
\tablecomments{
GMP galaxy identification numbers are from the Coma 
catalog of Godwin, Metcalfe, \& Peach (1983), as are the 
Right Ascension, Declination, and the photographic $b$-band apparent galaxy 
magnitude ($m_b$) within an isophote $\mu_b=26.5$. 
The recessional velocities have come from 
Matkovi\'c \& Guzm\'an (2003, in prep), and the morphological type 
has been estimated by us. 
}
\end{deluxetable}

\clearpage
\begin{deluxetable}{lcccccccccccccc}
\footnotesize
\tablecaption{Structural Parameters. \label{tbl-2}}
\tablewidth{0pt}
\tablehead{
\colhead{GMP}   & \colhead{$\mu_{\rm 0,d}$}   & 
\colhead{$h$}             &  \colhead{$\mu_{\rm e,b}$}  &  
\colhead{$R_{\rm e,b}$}   &  \colhead{$n_{\rm b}$}      &  
\colhead{$\epsilon_b, \epsilon_d$}  &
\colhead{$M_{\rm nuc}$}    &  \colhead{{\it FWHM}}  &
\colhead{$M_{\rm b}$}     &  \colhead{$M_{\rm tot}$}      &
\colhead{$\mu_{\rm 0,b}$} &  \colhead{$<$$\mu$$>$$_{\rm e,b}$}&
\colhead{$\mu_{\rm e,tot}$}    &  \colhead{$R_{\rm e,tot}$}
\\
1 & 2 & 3 & 4 & 5 & 6 & 7 & 8 & 9 & 10 & 11 & 12 & 13 & 14 & 15
}
\startdata
2571  &...   &...  &22.98 &1.82 &1.30 &0.17, ...  &23.90 &0.12 &18.86 &18.85 &20.52 &22.16 &22.95 &1.80 \\
2585  &...   &...  &22.47 &2.82 &1.56 &0.48, ...  &...   &...  &17.31 &17.31 &19.43 &21.56 &22.47 &2.82 \\
2879  &...   &...  &22.29 &3.14 &2.01 &0.16, ...  &22.88 &...  &16.77 &16.77 &18.28 &21.25 &22.28 &3.12 \\
2955  &...   &...  &22.79 &1.80 &1.16 &0.56, ...  &...   &...  &18.75 &18.75 &20.61 &22.02 &22.79 &1.80 \\
2960  &20.34 &2.96 &20.63 &1.27 &1.56 &0.06, 0.51 &...   &...  &17.20 &15.68 &17.59 &19.72 &21.57 &3.78 \\
2983  &...   &...  &23.21 &1.79 &1.74 &0.14, ...  &23.91 &...  &18.99 &18.98 &19.78 &22.25 &23.19 &1.76 \\
2985  &...   &...  &22.23 &2.74 &1.39 &0.38, ...  &22.57 &0.12 &17.19 &17.19 &19.56 &21.38 &22.21 &2.71 \\
3018  &...   &...  &22.61 &1.98 &1.09 &0.03, ...  &22.24 &0.17 &18.39 &18.36 &20.58 &21.87 &22.55 &1.92 \\
3209  &...   &...  &22.58 &1.64 &1.26 &0.18, ...  &24.76 &...  &18.71 &18.70 &20.19 &21.77 &22.57 &1.63 \\
3292  &20.68 &2.12 &21.82 &1.23 &2.23 &0.04, 0.20 &22.17 &...  &18.28 &16.74 &17.33 &20.73 &21.97 &2.90 \\
3406  &...   &...  &22.20 &1.84 &1.59 &0.31, ...  &24.01 &...  &17.96 &17.96 &19.10 &21.28 &22.19 &1.83 \\
3438  &...   &...  &22.86 &2.33 &1.38 &0.05, ...  &23.47 &0.14 &18.17 &18.17 &20.22 &22.01 &22.84 &2.31 \\
3486  &21.49 &4.29 &20.23 &0.83 &1.59 &0.25, 0.10 &...   &...  &17.72 &16.06 &17.15 &19.31 &22.85 &5.44 \\
3625  &...   &...  &23.62 &2.54 &1.38 &0.17, ...  &24.47 &...  &18.75 &18.74 &20.97 &22.77 &23.60 &2.52 \\
3629  &...   &...  &23.22 &2.77 &1.29 &0.45, ...  &23.93 &0.12 &18.19 &18.18 &20.77 &22.40 &23.20 &2.75 \\
3645  &...   &...  &20.98 &1.28 &1.62 &0.24, ...  &21.96 &0.31 &17.52 &17.50 &17.82 &20.05 &20.94 &1.25 \\ 
3806  &...   &...  &22.99 &1.65 &1.07 &0.12, ...  &24.55 &0.15 &19.18 &19.17 &21.02 &22.26 &22.97 &1.63 \\
3856  &...   &...  &23.25 &2.58 &0.94 &0.41, ...  &23.61 &...  &18.53 &18.52 &21.56 &22.58 &23.24 &2.56 \\
\enddata
\tablecomments{Column 1: GMP galaxy identification number is from the Coma 
catalog of Godwin, Metcalfe, \& Peach (1983). 
Column 2 and 3: Central surface brightness and scale-length of the outer 
exponential component (when one was fitted).  Column 4, 5, and 6: S\'ersic 
parameters of the host galaxy (or bulge component if an outer exponential 
was fitted).  (All of these model parameters were obtained by fitting the 
intermediate-axis light-profile 
$R=\sqrt{ab}$, observed through the {\it HST} F606W filter.)  
Column 7: Ellipticity $\epsilon_b$ at $R_{e,b}$, 
and the ellipticity $\epsilon_d$ at the outermost modelled data point when an outer 
exponential component was detected. 
Column 8:  Magnitude of the nuclear (point-source or extended Gaussian) component. 
The {\it FWHM} (in arcsec) of the extended, nuclear 
Gaussian (before convolution with the Moffat PSF) is given in Column 9. 
Column 10 and 11: Magnitude of the S\'ersic component and the total galaxy 
(including the nuclear and outer exponential component(s), if detected). 
Column 12 and 13: Central surface brightness, and mean surface brightness 
within the effective radius $R_{e,b}$, of the S\'ersic component. 
Column 14 and 15: Surface brightness 
$\mu_{\rm e,tot}$ at the (total) galaxy half-light radius $R_{\rm e,tot}$. 
Assuming the Coma cluster is at a distance of 100 Mpc, and $H_0$=70 
km s$^{-1}$ Mpc$^{-1}$, 0.1$\arcsec$=47 pc. 
}
\end{deluxetable}

\end{document}